\documentclass[twocolumn,showpacs,preprintnumbers]{revtex4}
\usepackage{dcolumn}
\usepackage{bm}
\input epsf

\begin{document}


\preprint{gr-qc/0603013}

\title{Tachyonic Quintessential Inflation}

\author{V\'{\i}ctor H. C\'{a}rdenas}
\email{vcardenas[at]unab.cl}


\affiliation{Departamento de F\'{\i}sica y Matem\'atica,
Universidad Andres Bello, Los Fresnos 52, Vina del Mar, Chile}

\begin{abstract}
We study the possibility to construct an observationally viable
scenario where both early Inflation and the recently detected
accelerated expansion of the universe can be explained by using a
single scalar field associated with the Tachyon. The Reheating
phase becomes crucial to enable us to have a consistent cosmology
and also to get a second accelerated expansion period. A
discussion using an exponential potential is presented.
\end{abstract}


\maketitle

\section{Introduction}

The ultimate goal for a cosmological model is to make all the
observational data available consistent within its framework. In
this context, a model which try to describe the current
accelerated expansion must also be consistent with large scale
structure observations, cosmic background measurements and the
standard model of cosmology. In this paper I propose a model to
describe both the current cosmic accelerated expansion and also
the early phase known as inflation \cite{Inf} consistent with the
standard model of cosmology.

Models of this type have been proposed in the past and are usually
called ``Quintessential Inflationary" models \cite{QI}. They are
characterized by a single scalar field which evolve in a non
oscillatory potential. The field drives both accelerated phases by
rolling down the potential to an asymptotic flat region. The
contact with the standard model of cosmology is made by
introducing a reheating phase -- where almost all the particles in
the universe were created -- through a gravitational particle
production process \cite{GP}. The shape of the potential has to be
adjusted to ensure a kinetic dominated phase during which the
scalar field contribution becomes negligible and the structure
formation and nucleosynthesis processes works without any
interference. The advantage of using a single scalar field to
describe both accelerated phases is lost when we introduce an
ad-hoc potential, adjusting its shape. Actually, this procedure
leads to a problem of consistency with the amplitude of density
perturbations called the $\eta$ problem \cite{eta}.

In this paper I propose a similar model which describe both
periods of cosmic acceleration by a single scalar field which we
identify with the tachyon leading to what we called ``Tachyonic
Quintessential Inflation" model. As far as I know, the tachyon has
been used to study cosmology in different contexts; studying
inflation, dark energy and also as a candidate to model dark
matter and dark energy together. Here I present a distinct model;
a single framework to describe inflation and dark energy. The main
advantage of using the tachyon as the field driven both phases is
not only the economy of scalar fields used, but also that \emph{we
know} from first principles the shape of the scalar potential to
be used \cite{sen98, T&C}.

Tachyons and inflation have been studied in several papers
\cite{F&T, K&L, T&I}. The idea is very simple but some of them are
fine tuned and have a problem of consistency with the scale of
energy density fluctuations \cite{K&L}. We refer to section II for
a discussion about that. Tachyons and dark energy have been
considered  a couple of times \cite{NOD}. We can understand that because the
usual exit from tachyonic inflation leads to an asymptotic
equation of state $P=0$; describing dust or dark matter. The way
in which we can obtain dark energy from tachyons is discussed in
section III through the incorporation of a reheating phase. During
this period, that occurs after inflation, the tachyon energy
density decreases many orders of magnitude enabling both have a
consistent cosmology \cite{SW} and also to get the possibility of
a new accelerated expansion phase. The main contribution of our
paper is in section IV. There we describe the model and its
relation with the two previous sections. In particular we
concentrate in the analog model of Quintessential Inflation where
the scalar field here is played here by the Tachyon. I end the
paper with a summary.


\section{Tachyons and Inflation}
Let us consider the cosmological consequences of a matter
component given by the Tachyon through the Born-Infeld action
\begin{equation}
S=\int \sqrt{-g}d^{4}x\left[ \frac{R}{16\pi
G}-V(\phi)\sqrt{1-g^{\mu
\nu}\partial_{\mu}\phi\partial_{\nu}\phi}\right] \label{action}
\end{equation}
where $\phi$ is a scalar field associated with the Tachyon. We
have to stress here that this action has not been derived from
first principles. In fact, the way in which the Tachyon appears
coupled to gravity leads also to many other terms that are not in
our functional $S$. Actually it can be considered as a good
starting point in the study of Tachyons in curved space
\cite{T&C}. We work in a $3+1$ space-time with a flat FRW metric
\begin{equation}
ds^{2}=dt^{2}-a(t)^{2}d{\bf x}^{2} \label{FRW}
\end{equation}
where $a(t)$ is the scale factor. By computing the stress-energy
tensor we find that the Tachyon can be interpreted as a fluid with
energy density

\begin{equation}
\rho=\frac{V(\phi)}{\sqrt{1-\Pi^{2}}} \label{energy}
\end{equation}
and a negative density pressure
 \begin{equation}
p=-V(\phi)\sqrt{1-\Pi^{2}} \label{pressure}
\end{equation}
where $\Pi = \dot{\phi}$ and we have assumed that $\phi$ is a
homogeneous field. The equation of state of this component is
$p=(\Pi^{2}-1)\rho$ which for a constant potential value is
equivalent to the Chaplygin gas EoS.

The equation of motion for the scalar field is
\begin{equation}
\dot{\Pi}+(3H\Pi+\frac{V'}{V})(1-\Pi^{2})=0, \label{eomf}
\end{equation}
where $H=\dot{a}/a$ is the Hubble parameter and a prime means
$'=d/d\phi$. In general the Friedman equation can be written as
\begin{equation}
H^{2}=\frac{8\pi G}{3}[\rho_{\phi}+\rho_{m}] \label{feq}
\end{equation}
The shape of the tachyonic potential depends on the system under
consideration; for example, from bosonic string theory the
potential has a maximum at $\phi=0$, where that maximum $V=V_{0}$
is the tension of some unstable bosonic D-brane, and a local
minimum with $V=0$ usually at $\phi \rightarrow \infty$. In this
paper we consider a well motivated expression
\begin{equation}
V(\phi)=V_{0}\exp(-\phi /\phi_{0}) \label{pot}
\end{equation}
which has been used in Ref. \cite{F&T} and the parameter $V_{0}$
is proportional to a D3-brane tension. The fact that this tension
and also the 4D Planck mass can be expressed in terms of the
string parameters, leads to a consistency problem among CMBR
observations, gravity waves detection, and inflationary
perturbations \cite{K&L, T&I}. Although some potentials can
circumvent this problem, (see also \cite{YSP}) I choose to work with potential
(\ref{pot}) and describe the phenomenology associated with a model
consistent with observations.

Let us study Inflation in this context. From (\ref{eomf}) and the
Friedman equation (\ref{feq}) we can derive the equation

\begin{equation}
\frac{\ddot a}{a}=\frac{8\pi G V}{3\sqrt{1-\dot{\phi}^2}}\left(
1-\frac{3}{2}\Pi^2\right), \label{adosp}
\end{equation}
from which we conclude that accelerated expansion occurs if the
condition

\begin{equation}
\Pi^2<\frac{2}{3},\label{endinf}
\end{equation}
is satisfy. From Eqs. (\ref{energy}) and (\ref{pressure}) we find
that a cosmological constant equation of state regime is reached
if initially the scalar field kinetic energy is very small
\begin{equation}
\Pi^2\ll 1.\label{cond1}
\end{equation}
To keep holding these two conditions as much as possible, we need
to ensure a slow roll regime of the field. It means a period where
\begin{equation}
\dot{\Pi}\ll 3H\Pi (1-\Pi^2).\label{cond2}
\end{equation}
From (\ref{cond1}) and (\ref{cond2}) we find the slow roll
equations of the system
\begin{equation}
H^{2}\simeq \frac{8\pi G}{3}V(\phi), \label{feq2}
\end{equation}
\begin{equation}
3H\Pi+\frac{V'}{V}\simeq 0. \label{eomf2}
\end{equation}
From these two equations we can rewrite the condition
(\ref{cond1}) in terms of the scalar field potential $V(\phi)$
leading to the well known inequality \cite{F&T, T&I}
\begin{equation}
\frac{(V')^2}{V^3}\ll 24\pi G, \label{epsilon}
\end{equation}
and doing a similar work with the second condition (\ref{cond2})
we find
\begin{equation}
\frac{V''}{V'\sqrt{V}}\gg \sqrt{24\pi G}.
\end{equation}
We can also compute the number of e-foldings during Inflation.
From the equations we have derived we find
\begin{equation}
N(\phi)=\int Hdt=-\int_{\phi}^{\phi_{end}}\frac{H^2 V}{V'}d\phi,
\end{equation}
which lead us to the expression
\begin{equation}
N(\phi)=\frac{8 \pi \phi_{0}^2}{M_{p}^2}V_{0}
\left[e^{-\phi_{i}/\phi_{0}}- e^{-\phi_{e}/\phi_{0}} \right],
\label{enefi}
\end{equation}
where $\phi_{i}$ and $\phi_{e}$ are the values of the tachyon
field at the beginning and end of the inflationary phase. The
factor in front of the parenthesis defines the dimensionless
parameter $X_{0}$ introduced by Fairbairn and Tytgat \cite{F&T}.
To solve the horizon and flatness problem we need at least a
number of $65$ e-folds of inflation. Because after inflation the
inequality (\ref{epsilon}) saturates (which is equivalent to the
condition $\epsilon \simeq 1$) implies (up to factors)
$X_{0}\simeq \exp (\phi _{e}/\phi _{0})$, so we are safe to
neglect the second term in eq.(\ref{enefi}). This implies the
restriction
\begin{equation}
16\pi \frac{V_{0}\phi_{0}^{2}}{M_{p}^{2}} e^{-\phi_{i}/\phi_{0}} >
65. \label{nef}
\end{equation}
Neglecting the exponential, the inequality implies that this
factor follows as in Ref. \cite{F&T}. The inclusion of the
exponential, which depends on the value $\phi_{i}$, enable us to
both satisfy the restriction (\ref{nef}) and also the one we shall
find on $\phi_{0}$ at the end of section IV.


\section{Reheating the Universe}

In a universe filled only by tachyons, the evolution of the scale
factor after inflation ends asymptotically as $a(t)\sim t^{2/3}$,
implying that there is no room for a second period of accelerated
expansion \cite{sen98, T&C,T&DE}. A simple way to see that, is to
consider the scalar field equation (\ref{eomf}). From the
beginning of inflation $\Pi$ increases with time, $\dot{\Pi} >0$.
The condition for that is
\begin{equation}
3H\Pi < \frac{1}{\phi_{0}}, \label{reh1}
\end{equation}
where we have used (\ref{pot}) in (\ref{eomf}). Because $H$ is a
decreasing function of time, $\Pi$ increases slowly and $\phi_{0}$
is a constant, the tachyon field never stop rolling down its
potential towards its asymptotic value $\Pi\rightarrow +1$, which
leads to $p\sim 0$ from Eq. (\ref{pressure}). However, to make
contact with the standard model of cosmology, we need a radiation
dominated period after inflation, and after that a matter
dominated phase appropriated for the large scale structure
formation process. This immediately implies that we have to
consider a period of reheating \textit{after} inflation.

Reheating is the period where almost all particles in the universe
were created and it begins just at the end of the inflationary
era. In the original Quintessential Inflationary model \cite{QI}
the reheating proceed through gravitational particle production
\cite{GP}. In this context the inflaton (the tachyon in this work)
does not decay, but it continues rolling down the potential during
the radiation and matter dominated phases of the universe. The
main problem here is the efficiency of the process.

An alternative mechanism of reheating is the introduction of
another scalar field in the model called "curvaton" \cite{curv}.
In this framework, the reheating is produced by the decay of the
curvaton field, meanwhile another field drive inflation or the
current acceleration in the case of dark energy models \cite{ucv}.
Although the mechanism is more efficient, the all idea is at odds
with the quintessential inflationary picture: the use of a single
field to describe both regimes.

Another way to reheat the universe is considering the decay of the
inflaton itself. In order to model that, we need to coupled the
tachyon field with other fields (bosonic and fermionic). Actually,
this is the way in which most of the inflationary models makes the
transition to a radiation phase \cite{Inf}. The advantage of this
approach, is that the energy density of the tachyon can be reduced
through its transformation into particles, enable us to make the
model consistent with observations \cite{SW}, because the tachyon
energy density is too large to be compatible with cosmology. In
fact, to have a tachyon field $\phi$ relevant today, we must have
a density parameter $\Omega_{\phi} \simeq O(1)$. Following
\cite{SW}
\begin{equation}
\frac{V_{i}}{\phi_{i}^{4}} \sim \frac{\rho_{today}}{\phi_{today}^{3}
\phi_{i}} \sim \frac{100 \text{ Gev }  \Omega_{\phi 0}}{\phi_{i}}
\end{equation}
where we have assumed that $V_{i}$ is defined at the beginning of
the radiation dominated period. Using the exponential potential
(\ref{pot}), and assuming that $V_{0} \sim M_{p}^{4}$ with $\phi_{i}
\sim \text{ TeV }$ scale we find that $\phi_{i}/\phi_{0} \sim 23$
implying a fall in 10 orders of magnitude for $\rho_{\phi}$.

In this paper I will assume a simplest phenomenological coupling
between the tachyon and radiation, in the spirit of recent works
\cite{CFM,STW,HPZ}. Let us consider a phenomenological coupling
between the tachyon and radiation. By writing the field equation
of the tachyon (\ref{eomf}) in its fluid form, we get
\begin{equation}
\dot{\rho_{\phi}}+3H\Pi^{2}\rho_{\phi}=-\Gamma \Pi^{2}
\rho_{\phi}, \label{rofi}
\end{equation}
and the equation for the created relativistic particles
\begin{equation}
\dot{\rho_{m}}+ 4H \rho_{m} = +\Gamma \Pi^{2} \rho_{\phi}.
\label{rom}
\end{equation}
Here the time scale for particle production $\Gamma^{-1}$ has to
be much less than the expansion time scale $H^{-1}$, because we
are interested in an efficient reheating process. So we assume
that
\begin{equation}
\Gamma \gg H.
\label{cond3}
\end{equation}
During reheating, the equation of motion for the field looks as
\begin{equation}
\dot{\Pi}+(1-\Pi^{2})(3H\Pi+ \Gamma \Pi -\frac{1}{\phi_{0}})=0,
\label{modeqf}
\end{equation}
which generalized Eq. (\ref{eomf}). Because of (\ref{cond3}) we
assume also that $\Gamma \gg 1/\phi_{0}$. This assumption is
necessary because otherwise, the effects of the particle creation
we want to consider would be negligible.

Neglecting the expansion of the universe, the evolution of the
tachyon field follows in two stages during reheating; in the first
one, the energy density $\rho_{\phi}$ falls down towards a finite
value with a $\Pi$ field evolving to the asymptotic value very
close to zero. In the second stage, the field $\Pi$ reaches a
stable configuration where it keeps the constant value
\begin{equation}
\Pi_{c} \simeq \frac{1}{\Gamma \phi_{0}}\ll 1. \label{fic}
\end{equation}
In this case, the tachyon energy density decreases exponentially
with time
\begin{equation}\label{decay}
\rho_{\phi}=V_{0}e^{-\phi/\phi_{0}}. \label{rofidec}
\end{equation}
We effectively have an exponential time decay of the energy
density, because $\Pi$ is a constant implying that $\phi (t)
\propto t$. Once the expression for the energy density
$\rho_{\phi}$ has been found, we insert it into the equation for
the created particles (\ref{rom}). However, this is not so direct,
because the right hand side of (\ref{rom}) is proportional to
$\Pi^{2} \rho_{\phi}$. During the first stage of reheating, the
field $\Pi$ falls down to $\Pi_{c}$, doing the effect of the
particle creation process less efficient. Using this fact and Eq.
(\ref{energy}) we notice that we can model the decay of the energy
density, during both stages of reheating, as
\begin{equation}
\rho_{\phi} = M^{4} e^{-\alpha \Delta t}.\label{ansatz}
\end{equation}
Using this in (\ref{rom}) the energy density of the created
particles behaves approximately
\begin{equation}
\rho_{m}(t) \simeq \frac{3 \Gamma M^{4} t_{0}}{8} \left[
\frac{t}{t_{0}}e^{-\alpha \Delta
t}-\left(\frac{t_{0}}{t}\right)^{8/3}\right ]
\end{equation}
where we have assumed that the Hubble parameter behaves as $H
\simeq 1/2t$. This assumption is natural because after inflation
the tachyon field follows to a matter like dominated universe.
Here $t_{0}$ is the time for which $\rho_{m}(t_{0})=0$, and marks
the beginning of the particle creation process. From this solution
we see that the energy density of the created particles reaches a
maximum of $\Gamma M^{4}/\alpha $, and then starts to decrease
according to a radiation dominated solution. Then during reheating
the tachyon equation of state interpolates between $\Pi \sim
\sqrt{2/3}$, a matter dominated regime and a vacuum one where $\Pi
\sim \left( \Gamma \phi _{0}\right) ^{-1}\ll 1$. After reheating
the universe becomes dominated by radiation, meanwhile the tachyon
slowly evolves towards a dust like equation of state (EoS). So the
coupling suggested (\ref{rofi},\ref{rom}) leads to the correct EoS
for a radiation dominated universe. Because $\Pi(t)$ varies during
this process, the estimation for the maximum is slightly larger
than the exact value.

Using condition (\ref{cond3}) in (\ref{modeqf}) and assuming that
$\Delta t \gg \Gamma^{-1}$ we find approximately that
\begin{equation}
\Pi(t) \sim \Pi_{0}e^{-\Gamma \Delta t} + \frac{1}{\Gamma
\phi_{0}}
\end{equation}
from which we can write an expression for the tachyon energy
density
\begin{equation}
\rho_{\phi} \simeq \rho_{\phi}(t_{0}) \exp{(-\Pi_{0}^{2}\Gamma
\Delta t)}.
\end{equation}
This result implies that we can use a value $\alpha =
\Pi_{0}^{2}\Gamma $ in (\ref{ansatz}) to obtain the maximum of the
radiation energy density $ \rho_{m}^{(max)} \simeq M^{4}$. For
this reason, once we solve the problem of doing the tachyon field
$\phi$ consistent with observations (leading to a $\Omega_{\phi}
\sim O(1))$, we also solve the adjustment of the matter component
at the end of inflation to obtain a regular standard model of
cosmology.

Although the reheating mechanism presented here is inefficient
compared to that for a standard scalar field, it can be improved
considering the preheating scenario \cite{preheat}. It is a
suitable mechanism for a model where the scalar field can not
oscillate around the minimum of the potential. During preheating
the transfer of energy from the scalar field follows a parametric
resonance channel which makes the decay process very fast.

\section{The tachyon as dark energy}

Using the exponential potential (\ref{pot}), the tachyon field
equation of motion (\ref{eomf}) admits a transient accelerated
expansion solution; a configuration where $\phi \gg \phi_{0}$ and
$\Pi \ll 1$. In terms of the variables defined in \cite{CGST},
similar to those used in Ref.(\cite{AL}), a configuration where
\begin{equation}
\lambda \equiv -\frac{M_{p}V'(\phi)}{V^{3/2}}\leq 1
\end{equation}
admits a transient accelerated expansion followed by a
deceleration phase for $\lambda \gg 1$. In our case this condition
leads to the inequality $\exp{(\phi/\phi_{0})} \leq X_{0}$, which
is consistent with the considerations at the end of section II. As
we have said in the previous section, during reheating the field
$\Pi$ ends with a very small value (\ref{fic}) enabling us to get
the appropriated conditions for a second period of exponential
expansion. Although the Hubble parameter $H$ decreases during the
evolution, after reheating the universe reach the following stage
\begin{equation}
\Pi  < \frac{1}{3H\phi_{0}}
\end{equation}
implying that the field rolls towards $\Pi \rightarrow +1$. A
second period of exponential expansion is possible if (see eq.
(\ref{adosp}))
\begin{equation}
\rho_{\phi}\sim \rho_{m}
\end{equation}
and the $\Pi$ field must satisfies $\Pi < \sqrt{2/3}$. Because at
the end of reheating the tachyon energy density is many orders of
magnitude less than the matter energy density, the right
conditions for a second phase of accelerated expansion can only
takes place after the scale factor has growth $10$ orders of
magnitude; just the right order of magnitude needed to make the
tachyon relevant for cosmology. The key observation here is the
following; the tachyon energy density evolve always
\textit{slower} than those from radiation and matter. In fact,
after reheating the equation of motion for $\phi$ (\ref{eomf}) can
be written as
\begin{equation}
\dot{\rho_{\phi}}+3H \Pi^{2}\rho_{\phi}=0.
\end{equation}
Because after reheating $\Pi$ is very small, the field evolve
slower than matter and radiation. Once the radiation dominated
universe has begun, the tachyon equation of motion (\ref{eomf})
indicates that, although the slope of $\Pi$ is positive, and the
Hubble parameter $H$ decreases with time, the field $\Pi$ takes a
long time to get the asymptotic value $+1$. It can be see
evaluating the time it takes from the initial value $\Pi_{i}$ (the
value obtained after reheating) to a final arbitrary configuration
(for example $\sqrt{2/3}$, the one needed to get a second period
of accelerated expansion).

From (\ref{eomf}) the slope of $\Pi(t)$ change from a quasi linear
regime to an asymptotic one, just at the time when
\begin{equation}
3H\Pi \sim \frac{1}{\phi_{0}}.
\end{equation}
During the radiation and matter dominated phase, the Hubble
parameter $H \sim 1/t $, so the time scale for the tachyon to
reach a value of order one (or $\sqrt{2/3}$) is
\begin{equation}
\phi_{0} \sim H^{-1},
\end{equation}
that of the Hubble time scale, which is many orders of magnitude
larger than the reheating scale $\Gamma^{-1}$ (see eq. (\ref{fic})
and the discussion below eq. (\ref{cond3})). This is the result
that indicates the possibility to get a new phase of accelerated
expansion today. This time however, would be a transient phase
doing the model consistent with the successful standard model of
cosmology. The key ingredient is the occurrence of reheating; it
not only makes the model consistent with observations \cite{SW},
but also enable us to have a new period of accelerated expansion.

\section{Summary}

In this paper I have investigated the possibility of using the
tachyon as the field driven a quintessential inflationary model. I
have described the mechanism to get an inflationary phase and also
the process of reheating, necessary to joint it with the standard
model of cosmology. The reheating phase becomes crucial in order
to both make the tachyon consistent with cosmology and also
provide the mechanism to assure a second period of accelerated
expansion. I have also found that the recent episode of
acceleration is a transient one, after which the universe enters
into a matter dominated universe from which they never return. Its
important to stress also that the orders of magnitude needed to
decrease the tachyon energy density, to make it compatible with
the standard cosmology, coincides with the orders of magnitude
necessaries to keep the standard Cold Dark Matter model for
structure formation intact during the evolution, enabling the
tachyon becomes relevant today. The present model alleviates part
of the fine tuning in the usual QI model. The adjustment of the
scalar field potential slope to get enough inflation and then a
long kination phase is not necessary here. Using the tachyon, we
have the advantage of using a high energy motivated potential that
drives both regimes; inflation and the current acceleration.


\section*{Acknowledgments}

VHC want to thanks Ioav Waga and R. Herrera for useful
discussions. VHC was supported by DI-UNAB grant 05-04.


\end{document}